\documentclass[twocolumn,showpacs,preprintnumbers,amsmath,amssymb]{revtex4}

\usepackage{graphicx}
\usepackage{dcolumn}
\usepackage{bm}

\begin{document}

\preprint{IKP/TUD/Sav01/04}

\setlength{\footnotesep}{0.2cm}

\title{Parity assignments in $^{172,174}$Yb using polarized photons\\
and the K quantum number in rare earth nuclei}

\author{D. Savran} \email[{\it E-mail address:} savran@ikp.tu-darmstadt.de]{}
\author{S. M\"uller}
\author{A. Zilges}%
\affiliation{%
Institut f\"ur Kernphysik, Technische Universit\"at Darmstadt, 
Schlossgartenstrasse 9, D-64289 Darmstadt, Germany
}%

\author{M. Babilon} \email[Present address: Institut f\"ur Kernphysik, Technische Universit\"at Darmstadt, D-64289 Darmstadt, Germany]{}
\affiliation{
A.W. Wright Nuclear Structure Laboratory, Yale University, 
New Haven, Connecticut 06520, USA
}%

\author{M.W. Ahmed}
\author{J.H. Kelley}
\author{A. Tonchev}
\author{W. Tornow}
\author{H.R. Weller}
\affiliation{
Triangle Universities Nuclear Laboratory, Duke University, 
Durham, North Carolina 27708, USA
}%

\author{N. Pietralla}%
\affiliation{%
Nuclear Structure Laboratory, Department of Physics \& Astronomy, 
State University of New York at Stony Brook, Stony Brook, 
NY 11794-3800}%

\author{J. Li}
\author{I. V. Pinayev}
\author{Y. K. Wu}
\affiliation{
Free Electron Laser Laboratory, Department of Physics, Duke University, 
Durham, North Carolina 27708, USA
}%

\date{\today}

\begin{abstract}
The 100 \% polarized photon beam at the High Intensity 
$\gamma$-ray Source (HI$\gamma$S) at Duke University
 has been used to determine the parity 
of six dipole excitations between 2.9 and 3.6 MeV in 
the deformed nuclei $^{172,174}$Yb in photon scattering
($\vec{\gamma},\gamma'$) experiments. The measured parities 
are compared with previous assignments based on the K quantum number
that had been assigned in Nuclear Resonance 
Fluorescence (NRF) experiments by using the Alaga rules. 
A systematic survey of the relation between $\gamma$-decay 
branching ratios and parity quantum numbers is given for the rare 
earth nuclei.
\end{abstract}


\maketitle

Low-lying dipole excitations in heavy nuclei have been studied 
extensively using the Nuclear Resonance Fluorescence (NRF)
 or photon scattering method, which provides a 
model-independent way to determine excitation energies, spins, 
decay widths, decay branchings, and transition probabilities \cite{Knei96}. 
The parity of a nuclear state can be determined
by either scattering unpolarized $\gamma$-rays and
measuring polarization in the exit channel, or
using linearly polarized $\gamma$-ray beam and measuring
the azimuthal angular distribution of the scattered photons.
For deformed even-even nuclei the K quantum number of $J=1$ states 
can be assigned within the validity of the Alaga rules \cite{alag55} 
from the electromagnetic decay branching ratio  
\begin{align}
R = \frac{B(\Pi 1; 1^\pi_K \to 2^+_1)}
         {B(\Pi 1; 1^\pi_K \to 0^+_1)} &= 
\frac{\Gamma_{1}}{\Gamma_{0}} \cdot 
   \frac{E^{3}_{\gamma}(1^\pi_K \to 0^+_1)}
        {E^{3}_{\gamma}(1^\pi_K \to 2^+_1)} ~ \\ \nonumber &=
\left\{
\begin{array}{ll}
2 &{\sf for} ~K=0 \\
0.5 &{\sf for}~ K=1
\end{array}
\right. ~ ,
\end{align}
where $\Gamma_{1}$ and $\Gamma_{0}$ denote the
decay widths to the $2^{+}_{1}$ and $0^{+}_{1}$ levels, respectively,
and $E_{\gamma}(1^\pi_K \to 2^+_1)$ and $E_{\gamma}(1^\pi_K \to 0^+_1)$ 
correspond to the energies of these transitions.

In general, there is no relation between the $K$ quantum 
number and the parity of a $J=1$ excitation \cite{Bohr75}. However, 
restricting oneself to dipole excitations that carry the largest 
part of the excitation strength one selects collective modes for 
which certain selection rules may exist. 
Within realistic calculations for deformed nuclei in the framework 
of the interacting boson model (IBM) \cite{Iach87} with $s$-- and 
$d$-proton and neutron bosons ($sd$-IBM-2), where negative
parity states are not included, all $J^{\pi}=1^{+}$ 
levels have a branching ratio corresponding to $K=1$ (e.g. the
bandheads of the $K=0$ octupole vibrational band). 
This suggests that those states with $J=1$ and branching ratios
corresponding to $K=0$ have negative parity.
Positive parity has generally been assumed in previous works
for all $K=1$ excitations in the energy range of 
the $M1$ scissors mode for calculating  
the summed $B(M1)$ strength, if no direct parity assignments were 
available. 
This rule of thumb was supported by $\gamma$-ray polarization 
measurements analyzing 
Compton-scattering asymmetries of the NRF $\gamma$-ray lines in 
some deformed nuclei of the Nd to Er even-even isotopic sequences 
\cite{Knei96}. 
It was concluded that at least the strong dipole excitations in 
sufficiently axially-symmetrically deformed nuclei decay according to 
the Alaga rules for $\Delta K =1$ (0) for positive (negative) 
parity.

The $K$ quantum number is a good quantum number only in the case of 
axially symmetric deformation and the aforementioned correlation 
between the parity of a strong dipole excitation and its decay 
branching ratio to the ground band has been tested in these nuclei 
for which their axial symmetry was considered to be well established. 
Very recently, it has been proposed  \cite{Iach03} that heavy 
rare earth nuclei in the mass region $A \approx 170$ might be close 
to the critical point of an axially-symmetric--to--triaxial shape 
phase transition. 
For triaxial shapes the Alaga rules do not hold and 
in this case the assignment of positive parity from the branching ratio 
is lacking a basis. 
If the stable nuclei in the $A \approx 170$ mass region would 
indeed exhibit a more pronounced triaxiality than the lighter stable rare 
earth nuclei, then the previous compilations 
\cite{Piet95,Sche03} of the total scissors mode's 
$M1$ excitation strength using parity assignments on the basis of 
decay branching ratios might contain a systematic error for nuclei 
with mass numbers $A \approx 170$.

The interacting boson model in its proton-neutron 
version ($sd$-IBM-2) represents a simple and useful model for the 
description of the evolution of the quadrupole-collective structure of 
heavy nuclei and of proton-neutron mixed-symmetry states such as the 
scissors mode. 
The description of triaxial deformation in the framework of the IBM 
involves two alternate approaches that differ significantly for the 
description of mixed-symmetry states. 
Either one may include cubic terms of the quadrupole-quadrupole 
interaction in an $F$-spin symmetric Hamiltonian for generating 
triaxial eigenstates \cite{Isac99}, or one might 
consider a situation close to the dynamical symmetry limit SU(3)$^\star$ 
of the standard two-body IBM-2 with structural parameters of opposite sign, 
$\chi_\nu = -\chi_\pi$ \cite{Iach87}. 
The latter approach breaks $F$-spin symmetry \cite{Isac86} 
and could cause a 
significant amount of $F$-spin mixing into the low-energy states 
if the Majorana interaction were not too strong \cite{Barr94}. 
The properties of mixed-symmetry states with $F$-spin quantum number 
$F= F_{\rm max}-1$ are very sensitive to the strength of the 
Majorana interaction \cite{Hart87} and, thus, to the amount of 
$F$-spin mixing in the low-energy wave functions \cite{Rich94}. 
In fact, $F$-spin multiplets of states of neighboring nuclei, including 
Yb nuclei, have been observed for symmetric states with $F= F_{\rm max}$
\cite{Hart87} and the scissors mode with $F= F_{\rm max}-1$ \cite{Levi00}. 
These observations suggest that $F$-spin is not severely broken 
in the corresponding nuclei which may rule out an SU(3)$^\star$-like 
description even for the nuclei in the mass region $A\approx 170$. 
More information, particularly on the goodness of the $K$ quantum number 
for mixed-symmetry states of nuclei in this mass region, would be 
highly desirable for estimating the relevance of 
$F$-spin breaking descriptions of a possible triaxiality at 
$A\approx 170$. 

The measured summed dipole excitation strength in Yb isotopes attributed 
to the $1^+$ scissors mode on the basis of decay branching ratios in 
the energy region 2--4 MeV is in good agreement with the observations for 
neighboring nuclei if one assumes positive parity for all $K=1$ states 
\cite{zilg90}. 
In order to confirm these parity assignments and to extend the 
systematics between branching ratios and parities of dipole states in 
rare earth nuclei to the mass region $A\approx 170$, 
we performed a series of experiments for determining parity quantum 
numbers for some of the strongest dipole excitations of $^{172,174}$Yb.

\begin{figure}
\includegraphics[width=\columnwidth]{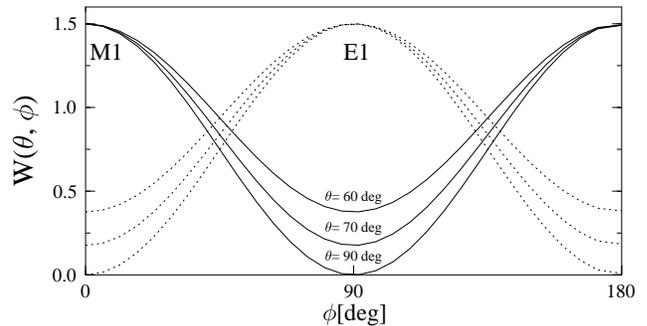}
\caption{\label{fig:ang} Azimuthal dipole angular distribution for resonant 
elastic photon scattering on a $0^+$ ground state for $M1$ (solid) 
and $E1$ (dotted) radiation, respectively, for different values of the 
polar scattering angle $\theta$. 
The difference between the two distributions at azimuthal angles
$\phi = 0^{\circ}$ and $\phi = 90^{\circ}$ is maximal for 
$\theta = 90^{\circ}$.}
\end{figure}

Parity quantum numbers of strongly dipole excited states can be 
assigned in NRF experiments by using a linearly polarized photon beam 
for excitation and by measuring the azimuthal angular distribution of 
the scattered photons about the polarization plane of the incident beam. 
For $0^{+} \stackrel{\vec{\gamma}}{\rightarrow} 1^{\pi_{1}} 
\stackrel{\gamma}{\rightarrow} 0^{+}$ 
elastic resonant photon scattering on the ground state of even-even 
nuclei due to dipole excitation, the resulting angular distribution 
is given by \cite{Fagg59} 
\begin{align}
W(\theta ,\phi ) = \frac{3}{2} + \frac{3}{4} 
\left(1-\cos^{2}\theta \right) \cdot 
\left[\pi_{1} \cos(2\phi ) -1 \right] ~ ,
\end{align}
where $\theta$ is the polar scattering angle with respect to the 
incident photon beam, $\phi $ is the azimuthal angle of 
the reaction plane with respect to the polarization 
plane of the incident $\gamma$-beam and $\pi_{1}$ is the parity 
quantum number of the excited state ($+1$ or $-1$). 
As shown in Fig. \ref{fig:ang},
the distribution for $E1$ transitions has a minimum at $\phi = 0^{\circ}$ 
and a maximum at $\phi = 90^{\circ}$, while the situation is vice versa for M1 transitions.
Therefore, it is sufficient to measure the angular 
distribution at $\phi =0^{\circ }$ and $\phi =90^{\circ }$ 
to determine the parity of a dipole state unambiguously.
For $\theta =90^{\circ}$ the minima of the distributions 
are zero and the analyzing power is maximal \cite{piet02}: 
\begin{eqnarray}
\Sigma ~=~ \frac{W(90^{\circ},0^{\circ})-W(90^{\circ},
90^{\circ})}{W(90^{\circ},0^{\circ})+W(90^{\circ},90^{\circ})}
\nonumber \\
=~ \pi_{1} ~=~ \left\{ 
\begin{array}{ll}
+1 &{\sf for} ~J=1^{+} \\
-1 &{\sf for} ~J=1^{-}
\end{array}
\right. ~ .
\end{eqnarray}

\begin{figure}
\includegraphics[width=\columnwidth]{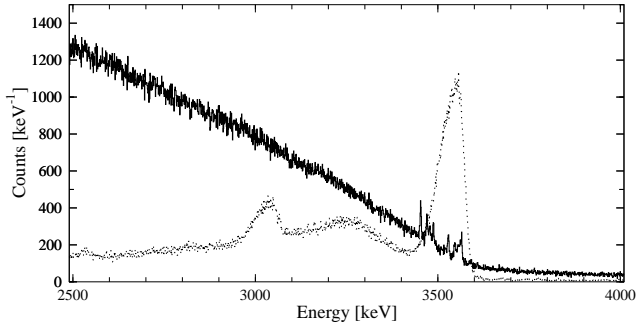}
\caption{\label{fig:beam} The energy distribution of the beam, which was obtained by reducing the 
photon flux and placing a germanium detector directly into the beam, is shown 
in the dashed spectrum. 
The events below 3450 keV are mainly due to the detector response, so 
the real energy distribution of the beam is given by the peak at 3550 keV,
which can also be seen in the measured $(\gamma ,\gamma ')$ spectrum (solid).
}
\end{figure}

The $\gamma$-beam \cite{Litv97} at the High Intensity $\gamma$-ray 
Source (HI$\gamma$S) at the Duke Free Electron Laser Laboratory 
combined with a set-up of four HPGe detectors 
has been proven to be useful for this type of experiment \cite{piet02}. 
The beam is generated 
by Laser Compton backscattering of photons from relativistic 
electrons and is quasi-mono-energetic. A free electron
laser is used, which is driven by the same electron beams. The backscattered
laser photons are boosted in energy by six orders of magnitude resulting
in $\gamma$ rays of several MeV with tunable energies. 
By adjusting the wavelength of the FEL and the energy of the electrons, 
one can select the desired energy of the $\gamma$ rays. 
In the present work $\gamma$ rays in the vicinity 
of 3 MeV had an energy spread, as defined mainly by the collimator opening angle,
of about 2~\% (FWHM) and the degree 
of horizontal linear polarization was better than 99~\%. 
We used a 45~g natural Yb$_{2}$O$_{3}$ target in order to be able to observe
excitations in different Yb isotopes simultaneously. 
The natural abundances of $^{172}$Yb and $^{174}$Yb 
are 21.9~\% and 31.8~\%, respectively. 
We measured at three mean energies of 2930 keV, 3005 keV and 3550 keV.
Fig. \ref{fig:beam} shows the summed spectra of all detectors
at a beam energy of 3550 keV. 
A coincidence with the electron pulses had been used to reduce 
the beam uncorrelated room background.
Peaks corresponding to elastic and inelastic photon scattering off the 
Yb nuclei are visible in the vicinity of the energy of the $\gamma$ ray 
beam. 
Below this energy, the spectrum consists only of continuous background 
mainly due to Compton scattering on various components of the 
experimental apparatus. 
The spectrum measured with a germanium detector positioned directly in the beam path 
is also included in the figure. The peak at 3550 keV exhibits the energy distribution of the beam.
The average intensity on target was about $5 \cdot 10^{5} ~ \gamma$ rays 
per second. The four Ge(HP) detectors were positioned 
at $\theta = 90^{\circ}$ and $\phi = 0^{\circ}, 90^{\circ}, 
180^{\circ}$ and $270^{\circ}$ at a distance of 5.5~cm from the target. 

\begin{figure}
\includegraphics[width=\columnwidth]{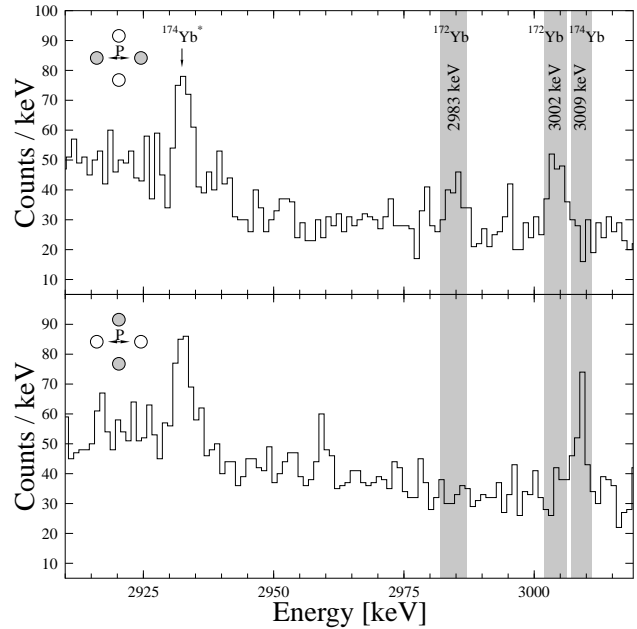}
\caption{\label{fig:spec} Photon scattering spectra obtained 
using a natural Yb target at $\theta =90^{\circ}$ and an beam energy of 3005 keV
parallel and perpendicular to the polarization plane. The peaks in the grey 
shaded areas correspond to known dipole transitions. 
Each peak is only visible in one of the spectra. 
The peak marked with an arrow results from an inelastic decay. }
\end{figure}

Fig. \ref{fig:spec} shows the energy spectra 
obtained with a beam energy of 3005 keV 
in the polarization plane of the beam and perpendicular 
to it. Three $J=1$ states were populated at
this beam energy. The dipole states at 
3009 keV and 3002 keV were known from previous experiments \cite{zilg90}.
The first peak only appears in the upper spectrum, yielding 
$J^{\pi}=1^{+}$, while the second is only observed in the lower 
spectrum, yielding $J^{\pi}=1^{-}$. In $^{174}$Yb, the inelastic 
decay of the 3009 keV level to the first $2_{1}^{+}$ state at 
76.5 keV is visible in both spectra, as one would expect 
from the nearly isotropic angular correlation for the 
$J=1 \rightarrow J=2$ transition in a $0 \rightarrow 1 \rightarrow 2$ 
cascade.
A new $J^{\pi}=1^{-}$ state at 2983 keV was also observed. The existence of this 
state was confirmed by reanalyzing data from previous
NRF experiments using non-polarized bremsstrahlung \cite{zilg90}.  

\begin{figure}
\includegraphics[width=\columnwidth]{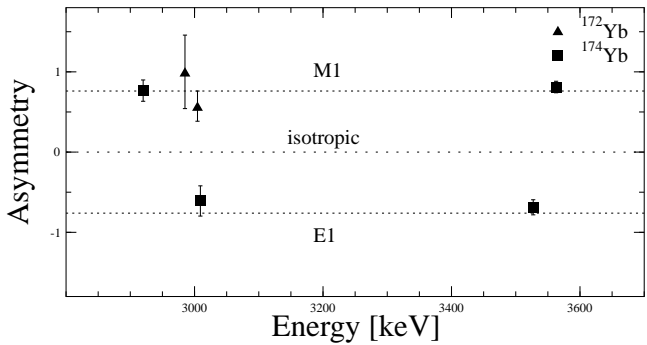}
\caption{\label{fig:asym} Experimental asymmetries 
determined in this experiment for the dipole states 
in $^{172,174}$Yb. The measured asymmetries are in good 
agreement with the expected values of $\epsilon = 0.76$ for M1 
transitions and $\epsilon = -0.76$ for E1 transitions } 
\end{figure}

The experimental asymmetry is given by 
\begin{align}
\epsilon = \frac{N_{\parallel} - N_{\perp}}{N_{\parallel} 
+ N_{\perp}} = q\cdot \Sigma ~,
\end{align}
where $N_{\parallel}$ and $N_{\perp}$ denote the efficiency-corrected 
peak areas in the spectra obtained by detectors located in the 
polarization plane of the incident photon beam or perpendicular to 
it. 
The experimental sensitivity $q$ of the setup is somewhat 
smaller than 1.0 as a result of the finite size of the target and 
the finite solid angles of the detectors. 
The asymmetries measured for the six excitations covered in our experiment
are summarized in Table \ref{tab:asym} and shown in Fig. \ref{fig:asym}.
We note that only states within the narrow energy range of the beam are
excited.
The calculated sensitivity for 
the present setup is $q = 0.761(5)$, which is in 
agreement with the observed asymmetries. For all cases, 
a parity assignment with a confidence level of more than 
four standard deviations is possible. All $J_K = 1_0$ states 
have negative parity, and all $J_K = 1_1$ states have positive parity. 
{\it I.e.}, for the strongest dipole excitations that carry the largest 
fraction of the respective total dipole strength the previous 
parity assignments done on the basis of decay branching ratios were 
correct and no discrepancy is found for the summed $B(M1)$ strength 
calculated in \cite{zilg90}. 

\begin{table}
\caption{\label{tab:asym} Measured asymmetries, 
assigned parities and branching ratios for the 
dipole excitations in $^{172,174}$Yb that were observed 
in this experiment.}
\begin{tabular}{ccccc}
\hline
\hline
\\[-0.3cm]
$E_{x} (keV)$& Isotope& $\epsilon$ &$J^{\pi}$ $(\hbar)$ & $R_{exp}$\footnote{taken from \cite{zilg90}} \\
\hline
\\[-0.3cm]
2920 & $^{174}$Yb & ~0.77(13)~  & ~$1^{+}$~ & ~0.44(8)~ \\
2983 & $^{172}$Yb  & 1.00(45)  & $1^{+}$ & 0.37(13) \\
3002 & $^{172}$Yb  & 0.57(19)  & $1^{+}$ & 0.55(11)\\
3009 & $^{174}$Yb  & -0.61(19) & $1^{-}$ & 2.75(52) \\
3527 & $^{174}$Yb  & -0.69(9)  & $1^{-}$ & 1.87(28)\\
3562 & $^{174}$Yb  & 0.81(7)   & $1^{+}\footnote{parity also measured in \cite{bohl84b}}$ & 0.50(10)\\

\hline
\hline
\end{tabular}
\end{table}

\begin{figure}
\includegraphics[width=\columnwidth]{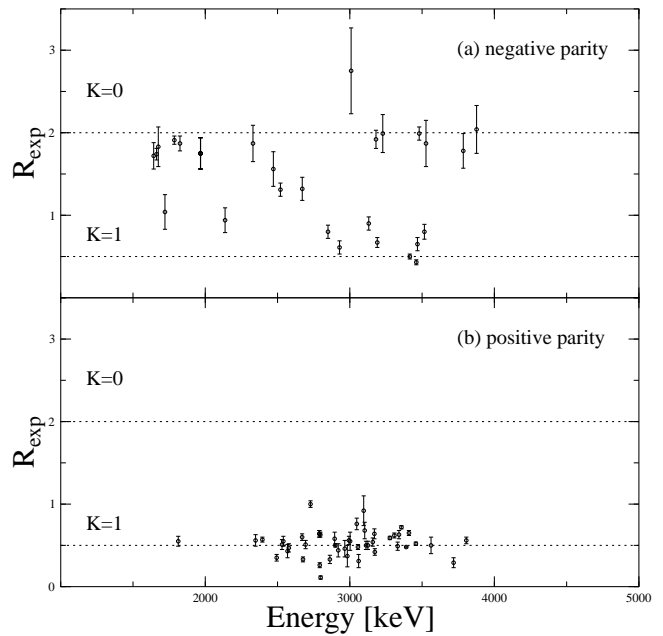}
\caption{\label{fig:k_pi} Measured branching ratios $R_{exp}$ 
for dipole states in different rare earth nuclei with negative parity 
(a) and positive parity (b). All states with $R_{exp} > 1$ 
have negative parity. }
\end{figure}

\begin{figure}
\includegraphics[width=\columnwidth]{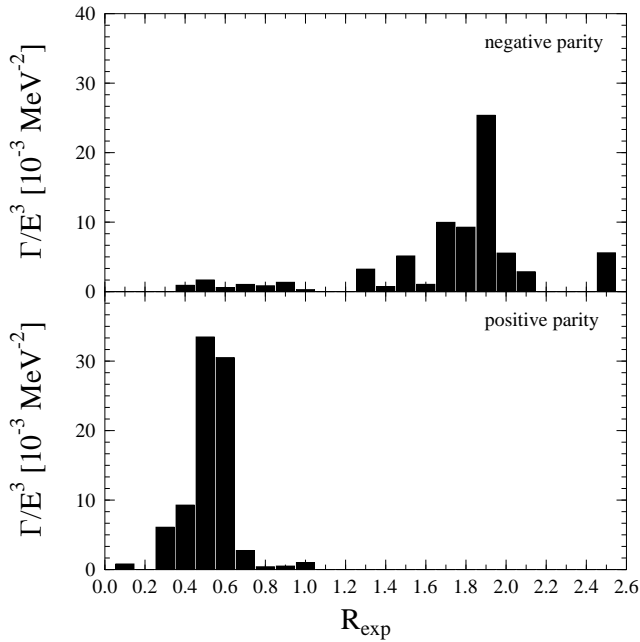}
\caption{\label{fig:summed} Summed dipole strength in bins of $\Delta R_{exp} = 0.1$. 
In case of $M1$ transitions the given values can be converted in $B(M1)$ strength by 
$B(M1)/\mu_{N}^{2}= 2.59 \cdot 10^{-1} \cdot \Gamma / E^{3}$. 
In case of $E1$ transitions by  $B(E1)/e^{2} fm^{4} = 2.87 \cdot 10^{-3} \cdot \Gamma / E^{3}$. 
}
\end{figure}

In the present work we are dealing with well-deformed rotational nuclei 
with $E(4_{1}^{+})/E(2_{1}^{+}) \ge 3.0$, where $K$ is expected to be 
a good quantum number.
The correlation between the branching ratio $R_{exp}$ and 
the parity in well deformed rare earth nuclei is shown in Fig.~\ref{fig:k_pi} for all  
$J=1$ states where both values are known. The additional data 
points are taken from previous NRF experiments 
on $^{150}$Nd \cite{Marg93}, $^{160}$Gd \cite{Frie94}, 
$^{162,164}$Dy \cite{Marg95}, $^{166,168,170}$Er \cite{Mase96} 
and $^{176}$Hf \cite{Sche03}.
All positive parity states have a branching ratio of $R_{exp} < 1$, 
and most are consistent with a $K=1$ assignment, which is in 
agreement with the predictions for the scissors mode.  
Therefore, the assignment of negative parity for $J_K=1_0$
states seems to be justified. But there are also negative parity states 
with $R_{exp}<1$, especially some with a branching ratio that is 
consistent with $K=1$. 
The $K$-quantum number does not provide enough information
to make a parity assignment of $J_K=1_1$ states. However, 
most of the comparatively strong excitations with 
$R_{exp} \approx 0.5$, which corresponds to $K=1$, are positive 
parity states. Fig. \ref{fig:summed} shows the summed dipole strength over all
states shown in Fig. \ref{fig:k_pi} in bins of $\Delta R_{exp} = 0.1$ interval. 
For better comparison the $\Gamma / E^{3}$ values
are shown, which differ only by a faktor of $2.87 \cdot 10^{-3}$ and 
$2.59 \cdot 10^{-1}$ from the corresponding $B(E 1)$ and $B(M1)$ values, 
respectively. The fraction of electric dipole strength in the region below
$R_{exp} < 1$ is allmost negligible compared to the magnetic dipole strength 
in this region. Therefore, the calculated summed $B(M1)$ strength reported in previous 
works \cite{Piet95,Sche03} seems to be an acceptable upper limit.

It should be emphasized, that due to the low sensitivity of Compton 
polarimetry, the parity is nearly exclusively known only for the 
strongest excitations. The correlation
between branching ratio and parity presented above may not be observed
in the case of weaker excitations.

Let us finally return to the Yb nuclides studied in this
work and address the goodness of the $K$ quantum number of
the scissors mode of $^{172,174}$Yb.
Our polarity measurement has not only proven the validity
of previous positive parity assignments to some strong $M1$
excitations around 3 MeV as discussed above, but it
supported the data on Yb nuclides used for previous systematic
investigations of the scissors mode's excitation energy and
strength.
With respect to excitation energy and decay pattern we consider
the identified strongest fragments of the scissors mode (see
$1^+$ states in Tab.I) as representatives for the mode.
All four strong $M1$ excitations show, within the error bars
of about 20\%, a decay branching ratio $R_{\rm exp} = 0.5$, i.e.,
the Alaga value for pure $K=1$.
Possible impact of triaxiality on mixed-symmetry states is hence
not observed at this level of accuracy.
In fact, the remarkably high excitation energy of the $2^+$
$\gamma$ vibrational state (1465.9 keV for $^{172}$Yb and 1634.0
keV for $^{174}$Yb) points at comparatively rigid axial symmetry
for these nuclei.
Quantitative conclusions along the same line have been drawn
earlier from a detailed band mixing analysis in Ref. \cite{Cast83}.
Thus, both the proton-neutron symmetric collective structures
and the scissors mode with mixed proton-neutron symmetry support
a view of $^{172,174}$Yb as deformed nuclei with pronounced
axial symmetry.

We thank U. Kneissl for discussions and M. Pentico, V. Rathbone and P. Wang for the support during the beamtime. 
D.S., S.M., M.B. and N.P. thank Duke University for its 
kind hospitality. 
This work was supported by the DFG (SFB 634 and Zi510/2-2),
 the DAAD (D/0247211), the NSF (INT-02-33276) and the U.S. DOE under 
Grant No. DE-FG02-97ER41033.

\bibliographystyle{/userx/users/ds/docs/bibtex/apsrev}
\bibliography{/userx/users/ds/docs/bibtex/english}

\end{document}